# LAMINATION AND LASER STRUCTURING FOR A DEP MICROWELL ARRAY

*E. Jung, D. Manessis, A. Neumann, Lars Bšttcher, T. Braun,*
*J. Bauer, H. Reichl,* **Fraunhofer IZM, Germany**
*B. Iafelice,* **University of Bologna, Italy**
*Federica Destro, Roberto Gambari,* **University of Ferrara, Italy**

**ABSTRACT**

Microtechnology becomes a versatile tool for biological and biomedical applications. Microwells have been established long but remained non-intelligent up to now. Merging new fabrication techniques and handling concepts with microelectronics enables to realize intelligent microwells suitable for future improved cancer treatment.

The described technology depicts the basis for the fabrication of electronically enhanced microwell. Thin aluminium sheets are structured by laser micro machining and laminated successively to obtain registration tolerances of the respective layers of <5μm.

The microwells lasermachined into the laminate are with 50…350μm diameter, allowing to contain individual cells within the microwell as well as provide access holes for the layer-to-layer contacting

A permeable membrane attached to the bottom of the microwell plate is used for fluid handling.

The individual process steps are described and results on the microstructuring as well as on biocompatibility of the materials are given.

## 1. INTRODUCTION

Cell handling for cell-to-cell interaction detection is currently done in configurations mimicking patch clamp techniques, allowing precise control of the cell position [1]. Some techniques employed here are

- Glass pipette (std. patch clamp protocol) [2]
- Laser based tweezers [3]
- Electrostatic planar cells [4]
- $SiO_2$ structured Micromanipulators [5]

Using either of these techniques, the cell manipulation and detection of cell behaviour under certain influences (e.g. cell-cell interaction) are separate tasks done with separate equipment.

Also, none of the techniques fully employs the power of merging modern microelectronic control, planar fabrication techniques and circuit fabrication.

The described approach implements these techniques and makes use of

- Biocompatible materials selection *
- Lamination technique for mass manufacturable layer stacks *
- Laser structuring for small hole (i.e. a geometric confinement) and large area (i.e. control circuitry routing) *

As well as

- Surface modification *
- Tailored metal/dielectric layers for a dielectrophoretic control in 3D of cellposition *
- Cell nutrition through a selectively permeable membrane *
- and Cell-Dispensing into the microwell

The aspects marked with an asterix are described in the paper in detail.

Part of the structure that is used for determining the cell-cell interaction is depicted in /Figure 1/

E. Jung, D. Manessis, A. Neumann, L. Bšttcher, T. Braun, J. Bauer, H. Reichl,
B. Iafelice, F. Destro, R. Gambari

# LAMINATION AND LASER STRUCTURING FOR A DEP MICROWELL ARRAY

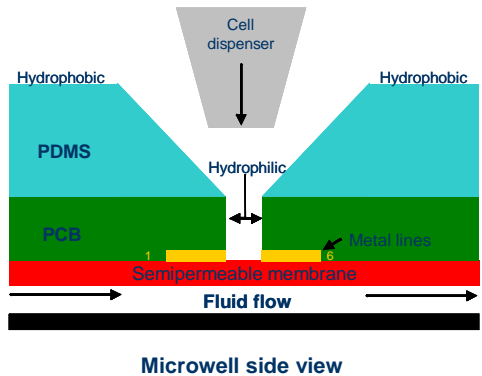

**Figure 1: Schematic of the cell-cell interaction analysis platform *Cochise***

A multilayer stack had to be created from non-cytotoxic materials and structured to accommodate per microwell the required volume for two biological cells (i.e. resulting in a 50..100µm diameter requirement for the microwell hole)

The pre-positioning of the cell into the microwell is done by a cell-jetting unit developed by CEA /[6]/ and facilitated by a funnel guiding the cell containing droplet into the microwell.

The main tasks described were therefore:

- Selection of suitable materials
- Development of processes capable of achieving the feature requirements
- Modification of the substrate surface to hydrophobic and hydrophilic, selectively
- Attachment of a semi-permeable membrane

## 2. PROOF OF CONCEPT REALIZATION

### 2.1 Selection of suitable materials

The selection of suitable materials had to take into account both the processability as well as the cytotoxicity of the candidate materials. Preferably, already known materials in either area were selected. For cost and availability reasons, Aluminium was selected before copper (cytotoxic) or gold (high degree of protein/cell adsorption) or platinum (high cost), because it provides excellent availability, low cost, inert and cytocompatible surface (native $Al_2O_3$) as well as good processing parameters.

Selection of the dielectric was done comparing polyimide, epoxy and acrylic. After cytocompatibility tests /according to e.g. [7]/ and process evaluation, the acrylic material PYRALUX was selected as adhesive dielectric, that also comes with liners of polyimide for better post-lamination control of the z-dimensions.

Cytotoxicity tests were performed on sterilized microwell samples containing the MUT (materials under test) in a defined fixture /Figure 2/

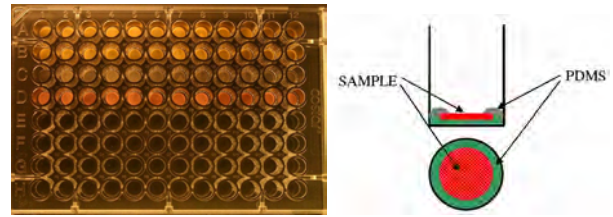

**Figure 2: Microwell sample for MUT and PDMS based fixture**

Table 1 summarizes the results of cytotoxicity tests performed.

| Mate | Effects on in vitro cell growth of LCL cells | |
|---|---|---|
| | *continous incubation (7 days)* | *pulse (60 min) and sub-culturing for 7 days* |
| Polyimide | no effects | no effects |
| Pyralux uncured | inhibition | no effects |
| Pyralux fullcured | no effects | no effects |
| PDMS | no effects | no effects |
| DIE attach film | inhibition | inhibition |
| Polyurethane | slow inhibition | no effects |
| Epoxy-B-Stage fullcured | inhibition | no effects |
| Au over Ni + ODT | no effects | no effects |
| Au over PD+ ODT | inhibition | no effects |
| Pd + ODT | no effects | no effects |
| Cu + ODT | inhibition | slow inhibition |
| Au over Ni | inhibition | no effects |
| Au over Pd | inhibition | slow inhibition |
| Pd | no effects | no effects |
| Cu | inhibition | inhibition |
| Al | no effects | no effects |
| Certonal FC-732 | no effects | no effects |

**Table 1: Cytotoxicity Tests on Material Samples**

### 2.2 Development of processes capable of achieving the feature requirements

The most straightforward approach to the realization was to use lamination techniques for layer stack creation and laser structuring for feature generation (pattern and holes)

*2.2.1 Layer lamination*

A Lauffer vacuum lamination press was used to laminate alternating layers of 15µm thick rolled Al-foil and 15µm thick Pyralux LF sheets onto each other. Using proper control of the material thickness, process parameters and selection of press suspensions (i.e.

polished stainless steel sheets) layers with minimum bow/warp and excellent layer thickness distribution could be achieved /Figure 3 & Figure 4/.

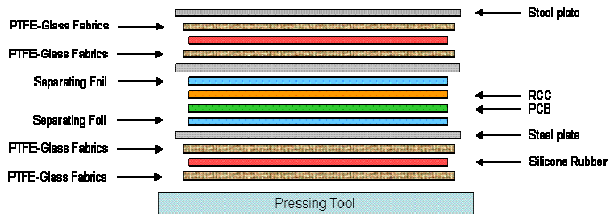

**Figure 3: Principle of a laminate stack for multi layer lamination (Tmax: 190°C, pmax: 20Bar)**

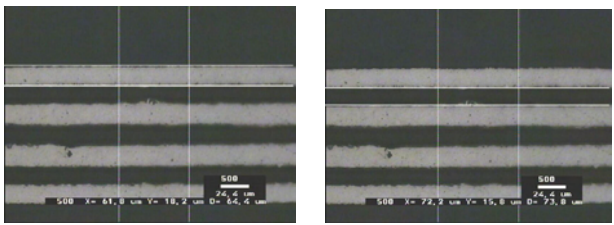

**Figure 4: Multi Layer Stack consisting of Al and Pyralux LF<sup>TM</sup>, alternating 15µm layers**

To note is here, that for the proof of concept a lamination of structured layers was necessary with a layer-to-layer alignment precision of 5..10µm – which cannot be achieved with the standard mechanical registration pins. For this, a layer by layer structuring was used, referencing the structuring laser according to b) towards the previously laminated metal layer, providing a nominal 2µm accuracy.

*2.2.2 Structuring*

In order to facilitate this and to maximize the flexibility (e.g. hole drilling and structuring), as Siemens Dematic Microline LaserDirectStructuring equipment was selected. Optimizing the process parameters, holes down to 50µm could be achieved through a 4x metal/dielectric layer stack. Hole quality however was only acceptable down to 80µm, still being in the specifications of the demonstrator requirements. On the other hand, also large holes were required for the technical studies – with the laser structuring it was possible to create the required hole diameters of 200….350µm as well /Figure 5a&b/.

Structuring of precision areas (electrode distance down to 25µm, only in ~0.2x0.2mm area) and large scale de-metallized areas (feature size ~75µm over several tens of cm$^2$ of area) was handled by a combination of resist protected structuring using Al-etchant and post-structuring using laser ablation cuts.

However, with the minimally achieved insulative trenches of 15µm a relative high amount of material build-up was detected, that could be minimized to a tolerable level by temporary sacrificial protective coating and adjusted cutting parameters to a 30µm trench size, acceptable for the current demonstrator specifications /Figure 6/.

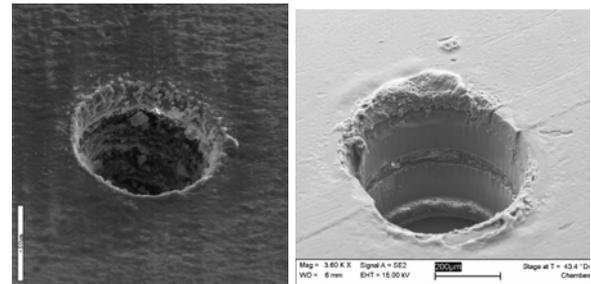

**Figure 5:    a                    b
Laser drilled hole in a 4x M/D layer stack
(left: 80µm diameter, right 300µm diameter)**

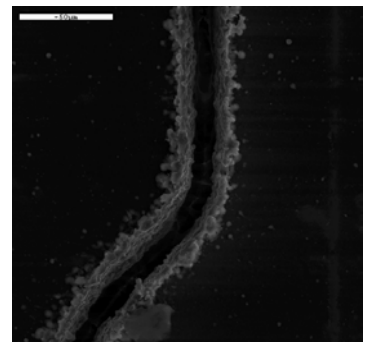

**Figure 6: Features cut into the top Al layer providing 15µm insulative trench**

In order to reach the final specifications of the target demonstrator, still some optimization is to be done.

**2.3 Modification of the substrate surface to hydrophobic and hydrophilic**

The inner part of the microwell hole should be hydrophilic, the outer surfaces of the substrate should provide hydrophobic properties to facilitate the trapping of a fluidic droplet containing the cells under investigation.

While the final structure is designed to feature a funnel like silicone structure /q.v. Figure 1/, for the first proof of concept the top surface was coated with surfactants to hydrophobize it.

The surfactants selected as candidate materials were tried in the cytotoxicity tests similar to the main materials used for the layer structure and from the candidates, two surfactants, octadecanethiol (ODT) and a fluoropolymer Certonal F372TM could be singled out as hydrophilizing resp. -phobizing agent.

The modification of the surface energy of the target surfaces (Al and Pyralux™ acrylate) was evaluated by contact angle measurement using a Dataphysics angiometer and the results are depicted in /Figure 8/.

The fluoropolymer coating provides the lowest surface energy with lowest polar contribution and therefore is the preferred surfactant for the modification. ODT on the other hand allows a lower contact angle, eg. better wetting of the surfaces and is therefore a candidate for hydrophilizing the used materials.

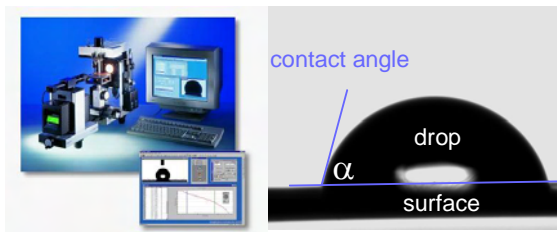

**Figure 7: Surface angle measurement technique by contact angle determination**

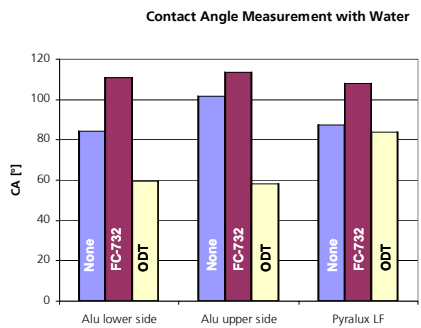

**Figure 8: Results indicating the effect of flouroacrylate FC732 and ODT**

### 2.4. Attachment of a semi-permeable membrane

To ensure both the nourishment of the cell in the microwell as well as removal of the metabolic products, the microwell is closed on the bottom by a semipermeable membrane.

Currently, selection of cellulose acetate CA and flouropolymer PTFE membranes with poresizes <5µm, manufactured by Millipore, is being performed for optimum results. The assembly process utilizes a pressure sensitive adhesive tape (PSAT), which is not affected by water immersion. This PSAT with a protective mylar backing is mounted prior the laser hole drilling process to the substrate. After the hole manufacturing & top surface modification, the protective backing is removed and the membrane is cold-laminated to the PSA and the process is completed /Figure 9/.

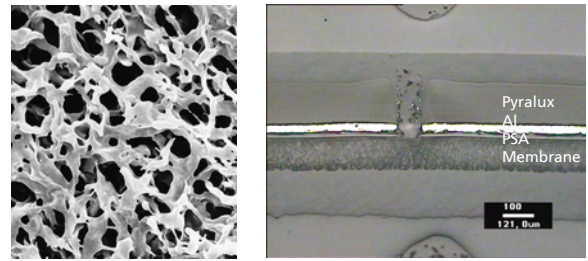

**Figure 9: Pore structure of the CA membrane, cross section of the assembled membrane on the bottom of the micro well array (80µm hole)**

### 3. *PROOF OF CONCEPT* FOR DEMONSTRATORS

The project demonstrator of Cochise will include all described process steps to achieve a structure depicted in /Figure 10/.

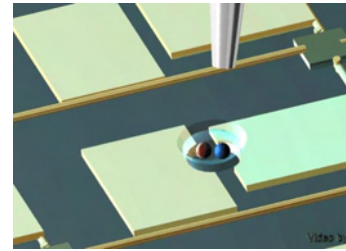

**Figure 10: Schematic of cell-cell interaction with the cochise platform /artwork done by UniBo/**

Targeting this goal, first implementation trials were carried out with the described process to verify i.e. the suitability of the impedance based control of the fluidic level inside the hole. For these trials, the schematic in /Figure 11/ depicts one of the measurement set-ups required to verify the control sequence.

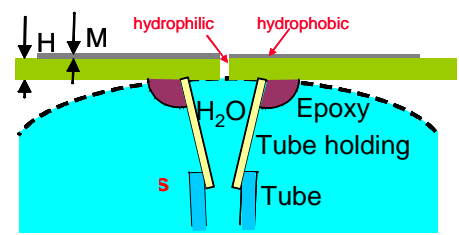

**Figure 11: Feasibility structure - Fluidic level determination by impedance measurement (H: Dielectric Thickness, M: Metal Thickness)**

E. Jung, D. Manessis, A. Neumann, L. Bšttcher, T. Braun, J. Bauer, H. Reichl,
B. Iafelice, F. Destro, R. Gambari

# LAMINATION AND LASER STRUCTURING FOR A DEP MICROWELL ARRAY

The measurement results of the fabricated feasibility studies indicate, that the concept is viable. However, it was found, that e.g. the roughness of the surface around the holes has a significant influence on the wetting and thereby on the measurement of e.g. the fluidic level.

Also, the variation of the electrode spacing will influence the level of control exerted on the position precision of cells/microbeads and is therefore critical.

With the laser tool, however, the actual position of the insulative cuts can be adapted to the actually achieved structures on the laminated stack, compensating for any warpage or distortion of the circuitry image.

## 4. FUTURE WORK

The results of the feasibility structures are promising and enable to fabricate the project demonstrator. However, process control using the laser drilling and structuring tool still needs optimization to achieve a high yield with burr-free exit holes/rims. Also, the necessity of a hydrophobic funnel may prove challenging because of the required alignment accuracy of >5µm and will be adressed using self aligned techniques.

## 5. CONCLUSION

A process flow to create a PCB like multi layer structure with biocompatible/non-cytotoxic material has been defined by using vacuum lamination, structured Al etching and laser direct structuring. Both small structures (50µm holes, 15µm insulative cuts as well as 350µm holes, 150µm pitched routing structures) have been created by combining the used processes to a defined flow. The stack lamination process was controlled to realize 1:1 aspect ratio metal:dielectric structures down to 15µm layer:layer thickness.

Hydrophobizing/hydrophilizing surfactants have been evaluated to selectively modify the exposed surfaces of the structure. A membrane attachment process has been realized, allowing to nourish cells confined in the microwells and simultaneously removing their metabolites.

## 6. ACKNOWLEDGEMENTS

The authors would like to acknowledge the support of the European Commission in the Cochise Project, EC contract 034534 as part of the FP6 framework program.

University of Bologna, Prof. R. Guerrieri´s working group, was in charge of the concept development, the design and simulation of the demonstrator and feasibility structures. The work is thankfully acknowledged.